\pdfoutput=1
\documentclass[aps,prl,twocolumn,longbibliography,floatfix,superscriptaddress]{revtex4-2}

\usepackage{babel}
\usepackage{amsmath}
\usepackage{chngcntr}
\usepackage{amssymb}
\usepackage{graphicx}
\usepackage{physics}
\usepackage[colorlinks=true,linkcolor=blue,urlcolor=blue,citecolor=blue]{hyperref}
\usepackage{comment}

\usepackage{epstopdf}

\usepackage{xcolor}

\usepackage[multiple]{footmisc}

\newcommand{\sA}{\textcolor[rgb]{0.917647 ,0.196078 ,0.137255}{A}}
\newcommand{\sB}{\textcolor[rgb]{0.215686 ,0.490196 ,0.133333}{B}}
\newcommand{\sC}{\textcolor[rgb]{0.458824  0.101961  0.486275}{C}}
\begin{document}

\title{Topological Devil's staircase in a constrained kagome Ising antiferromagnet}

\author{Afonso Rufino}
\email{afonso.dossantosrufino@epfl.ch}
\affiliation{Institute of Physics, Ecole Polytechnique F\'ed\'erale de Lausanne (EPFL), CH-1015 Lausanne, Switzerland}

\author{Samuel Nyckees}
\affiliation{Institute of Physics, Ecole Polytechnique F\'ed\'erale de Lausanne (EPFL), CH-1015 Lausanne, Switzerland}
\author{Jeanne Colbois}
\affiliation{Univ. Grenoble Alpes, CNRS, Grenoble INP, Institut NEEL, 38000 Grenoble, France}
\author{Fr\'ed\'eric Mila}
\affiliation{Institute of Physics, Ecole Polytechnique F\'ed\'erale de Lausanne (EPFL), CH-1015 Lausanne, Switzerland}

\date{\today}
\begin{abstract} 
We show that the constrained Ising model on the kagome lattice with infinite first and third neighbor couplings undergoes an infinite series of thermal first-order transitions at which, as in the Kasteleyn transition, linear defects of infinite length condense. However, their density undergoes abrupt jumps because of the peculiar structure of the low temperature phase, which is only partially ordered and hosts a finite density of zero-energy domain walls. The number of linear defects between consecutive zero-energy domain walls is quantized to integer values, leading to a devil's staircase of topological origin. By contrast to the devil's staircase of the ANNNI and related models, the wave-vector is \emph{not} fixed to commensurate values inside each phase. 
\end{abstract}
\maketitle

\noindent\textbf{\textit{Introduction.}}
Antiferromagnetic Ising models with farther-neighbor or frustrated interactions host a wide range of possible exotic phase transitions beyond the expected first-order or Ising universality classes~\cite{lacroix_introduction_2011,moller_magnetic_2009, chern_magnetic_2012, wannier_antiferromagnetism_1950, stephenson_isingmodel_1964,anghinolfi_thermodynamic_2015}.

Two famous examples of unusual phase transitions are particularly relevant here. The first is the Kasteleyn transition, originally described by Kasteleyn in 1963~\cite{kasteleyn_dimer_1963} in the context of dimer models~\cite{kasteleyn_statistics_1961, temperley_dimer_1961}. Also known as the Pokrovsky-Talapov transition, it occurs in constrained models~\cite{ pokrovsky_ground_1979, bhattacharjee_critical_1983, nagle_dimer_1989} where the only defects that respect the constraints span the whole lattice; thus, their energy grows linearly with the system size. As a consequence, they are completely absent up to the temperature at which their free energy becomes negative due to the entropic contribution. 

In frustrated models, the Kasteleyn transition may appear 
when constraints are imposed by infinite coupling constants. It is well known to play a key role in kagome ice~\cite{moessner_theory_2003,fennell_pinch_2007, turrini_tunable_2022}, as well as in spin ice under a [100] field~\cite{powell_classical_2008, jaubert_kasteleyn_2009, jaubert_three-dimensional_2008, szabo_fragmented_2022, alexanian_exploring_2024}. A case of direct relevance for the present study is the antiferromagnetic Ising model on the triangular lattice with second neighbor coupling~\cite{smerald_topological_2016}. The ground state consists of alternating lines of up- and down spins. It breaks the $\mathbb{Z}_3$ rotational lattice symmetry and the $\mathbb{Z}_2$ spin-flip symmetry. 
If a fifth-neighbor coupling is included,
and in the constrained limit of infinite first-neighbor coupling, the $\mathbb{Z}_2$ symmetry is first restored through a Kasteleyn transition before the $\mathbb{Z}_3$ symmetry is restored through a first-order transition.

The second important example is the devil's staircase, observed in the anisotropic next-nearest neighbor Ising (ANNNI) and related models, where an infinite series of commensurate steps is stabilized by thermal fluctuations instead of a floating incommensurate phase with a temperature-dependent wave-vector~\cite{bak_one-dimensional_1982, bak_commensurate_1982,fisher_infinitely_1980}.

In this Letter, we study the antiferromagnetic Ising model on the kagome lattice with up to third neighbor couplings (Eq.~\ref{eq:ham}) in a phase that breaks the rotational $\mathbb{Z}_3$ symmetry and the $\mathbb{Z}_2$ spin-flip symmetry. In the appropriate constrained limit of infinite first and third neighbor couplings, where the only allowed defects are system-spanning, we show that the transition is a highly exotic form of a Kasteleyn mechanism giving rise to a topological devil's staircase\cite{introNote1,introNote2}. Although linear defects are completely suppressed at low temperature, as in the Kasteleyn transition, their density, instead of growing as $(T-T_c)^{1/2}$, grows through an infinite and increasingly dense sequence of jumps. By contrast to the standard devil's staircase, this behavior is not related to commensurate steps, but is controlled by an integer index of topological origin, the ratio of the densities of linear defects and of zero-energy domain walls.

\begin{figure*}
    \centering
    \includegraphics[width=\textwidth]{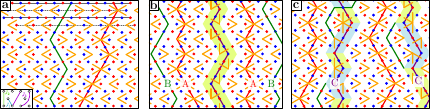}
    \caption{Spin configurations of the kagome lattice Ising antiferromagnet Eq.~\eqref{eq:ham} are represented by blue and red dots. Orange segments show an alternative representation.  (a) Main family of strings ground states : domains of left- and right-pointing orange arrows are delimited by \sA{} and \sB{} strings. The first dense and sparse rows which make up the Kagome lattice are highlighted as solid or dashed gray lines. Inset: labeling of the isotropic Ising couplings. (b) \emph{Zero-energy} double domain walls (DDWs) correspond to arrows rotated all in the same direction and are entropically suppressed in the thermodynamic limit. A thick purple line highlights the center of each DDW. (c) \sC{}-string excitations in the first plateau are characterized by different orientations of the arrows inside the DDWs, which introduce an energy cost. The density of \sC{}-strings ($n_C$) is equal to the density of ferromagnetic nearest-neighbors on a horizontal dense row.}
    \label{fig:1}
\end{figure*}

\noindent\textbf{\textit{Model and partially ordered ground states.}} 
We focus on the properties of the constrained model that corresponds to $J_1$ and $J_3$ infinite in the Hamiltonian:
\begin{equation}
E= J_1 \sum_{\langle i,j \rangle_1}  \sigma_i \sigma_j+J_2  \sum_{\langle i,j \rangle_2}  \sigma_i \sigma_j+J_3  \sum_{\langle i,j \rangle_3}  \sigma_i \sigma_j
\label{eq:ham}
\end{equation}
where $\langle i,j \rangle_{n}$ denote pairs of $n^{\mathrm{th}}$ nearest-neighbors on the kagome lattice, and $\sigma_i=\pm 1$. In contrast to already extensively studied models where the third-neighbor interaction is only across the hexagon~\cite{balents_fractionalization_2002-1, messio_kagome_2012-1, bernu_exchange_2013, gong_emergent_2014, gong_global_2015-1,rossi_schwinger_2023} or along the nearest-neighbor bonds~\cite{wolf_ising_1988,buessen_competing_2016, mizoguchi_clustering_2017, mizoguchi_magnetic_2018-1, grison_emergent_2020,colbois_artificial_2021,lugan_schwinger_2022} ~\footnote{A notable case is the SM of Ref.~\protect{\cite{andrade_partial_2024}}, which studies the $J_2 = 0$ limit with both 3rd neighbor couplings being non-zero.}, here, the third-neighbor interaction is set to the same value in both directions, forming triangular sublattices (Fig.~\ref{fig:1}a). 
For finite $J_1$ and $J_3$, the range $0<J_2\leq J_3< J_1$ corresponds to a single phase of ground state energy per site $E/N=-2J_1/3+2J_2/3-J_3/3$, referred to as the {\it strings} phase in Ref.~\cite{colbois_partial_2022}. In that phase, the macroscopic degeneracy of the nearest-neighbor model is only partially lifted, yielding a residual entropy equal to 1/3 of that of triangular lattice antiferromagnet.

This entropy is compatible with a simple family of ground states, illustrated in Fig.~\ref{fig:1}(a). Seeing the kagome lattice as a stacking of dense and sparse rows, the dense rows form a rectangular lattice. Consider a configuration on this rectangular lattice where the order is antiferromagnetic in the direction of the dense rows and ferromagnetic in the perpendicular direction. The energy minimum is reached if the remaining spins, which live on a $J_3$ triangular sub-lattice, are in the ground state of that model, with one frustrated bond on each $J_3$ triangle. Since the number of these spins is 1/3 of the total number of spins, the residual entropy of this family is 1/3 of that of the triangular lattice.
To visualize these states, it is convenient to draw dual-lattice bonds through the unsatisfied $J_1$ bonds. In the ground state, these dual bonds must not appear isolated; instead, they must belong to a pair of bonds meeting at 60º in the center of a hexagon, forming an arrow. If we define domains as contiguous regions where all arrows point in the same direction, the partially ordered family can be constructed by partitioning the lattice into domains with antiparallel directions~\footnote{This is similar in spirit to the construction of the triangular lattice Ising antiferromagnetic ground states starting from the stripe order and introducing double domain walls, see e.g.~\cite{smerald_topological_2016,smerald_spin-liquid_2018} and references therein.}. The boundaries between the uniform regions form domain walls living on the bonds of the triangular lattice and connecting the centers of the hexagons. We call these domain walls \sA{} or \sB{} lines if they are surrounded by converging or diverging arrows, respectively (see~\ref{sec:mapping}), or equivalently if they go through crossed or empty hexagons. In order to minimize $J_3$ interactions, the \sA{} and \sB{} lines may never propagate in parallel to the arrow directions or intersect, which forces them to form system-spanning zero-energy domain walls. Note that the number of \sA{} and \sB{} lines may at most differ by one since every domain must be bounded by an \sA{} line on one side and a \sB{} line on the other. In the thermodynamic limit, their densities must therefore be equal.

In this partially ordered family of states, the trajectories of \sA{} and \sB{} lines determine the spin configuration on the disordered $J_3$ triangular sub-lattice, where they separate antiferromagnetically aligned spins in the same row. Since adjacent spins on sparse rows have a $2/3$ probability of being antiferromagnetically aligned, the average density of \sA{} and \sB{} strings in the thermodynamic limit is $1/3$.\par
\begin{figure}
    \centering
    \includegraphics[width=\linewidth]{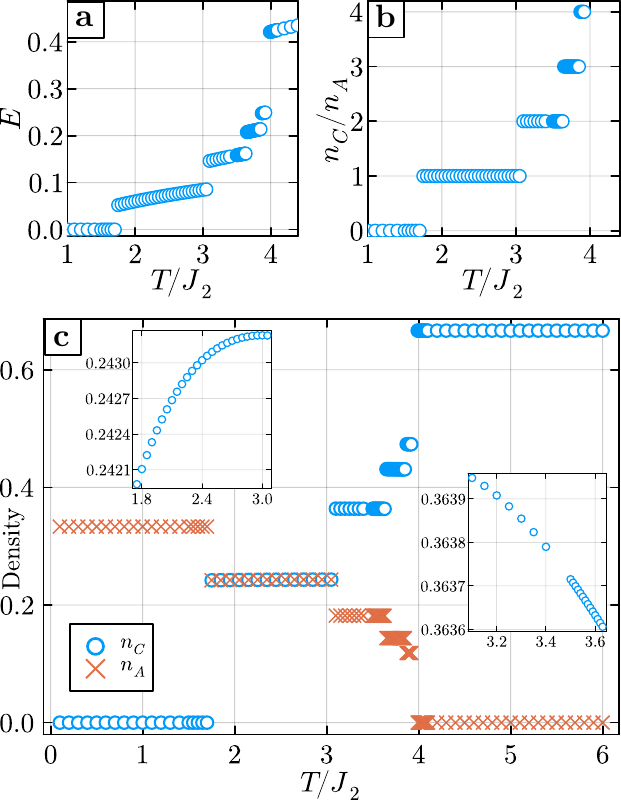}
    \caption{Numerical results from CTMRG for $\chi = 100$ suggest a topological devil's staircase. Only temperatures where the results are converged in bond dimension are shown (for more details on the bond-dimension dependence of CTMRG results, see \cite{sm}). (a) Energy. (b) Ratio of the strings density, revealing well-defined, integer-valued plateaus. (c) Density of \sA{} and \sC{} strings, related to the $\mathbb{Z}_3$ order parameter (see~\ref{sec:oderparameter}). A careful inspection reveals a continuous temperature dependence in the apparent plateaus of $n_C$, as shown in the two insets.}
    \label{fig:2}
\end{figure}

Interestingly, there is another family of zero energy extended defects, namely double domain walls (DDW) formed by replacing a \sB{} line with any number of arrows tilted by 60 or 120 degrees, as shown in Fig.~\ref{fig:1}(b). Such a DDW creates a defect in the antiferromagnetic order of the dense rows, and if a finite density of them (with random positions) was present in the ground state the $\mathbb{Z}_2$ symmetry would not be broken. These DDWs may meander in the same way that \sA{} and \sB{} lines do, but they have a thickness of one lattice unit. Thus, the entropy of configurations with one DDW is equivalent to the entropy of configurations with only \sA{} and \sB{} lines in a smaller system where one column has been removed. Since entropy is extensive, removing a column has an entropic cost linear in system size, and therefore DDWs are completely suppressed in the thermodynamic limit, inducing an entropy-driven \emph{partial} order~\cite{saglam_entropy-driven_2022,frenkel_order_2015}. This is in agreement with the numerical observation \cite{colbois_partial_2022} that the residual entropy is exactly that of the simple family of states with only \sA{} and \sB{} lines described above. In addition, there are also ground states which cannot be described by replacing \sB{} lines with DDW's. Their number grows sub-extensively with system size, and therefore they also do not contribute to macroscopic properties of the ground state~\cite{sm}.

\noindent\textbf{\textit{Excitations in the constrained limit.}}  
Let us discuss more specifically the constrained limit of infinite $J_1$ and $J_3$. In that limit, the energy of the triangles formed by $J_1$ and $J_3$ interactions is minimized, leading to the local configurations listed in FIG. \ref{fig:gs_tiles}. In this constrained model, the only possible excitations at finite temperature correspond to deviations from the $J_2$ ground state rule, which imposes a single bond to be unsatisfied in each $J_2$ triangle. Such excitations are only possible if a DDW is present (DDW which, in the constrained limit, must be system spanning; for a proof, see \cite{sm}). They are rooted in the development of domains with changes of the orientation of the arrows inside the DDW, possibly accompanied by portions of \sB{}-lines which decorate the DDW (see panel (c) of Fig.~\ref{fig:1}). From now on, these DDW with possible changes of orientation and decorations will be called \sC{}-lines. The energy cost of such \sC{}-lines comes from the presence of two types of local defects: (i) hexagons with two arrows that form a ``K" instead of a cross, with energy $4J_2/3$, and (ii) hexagons with three arrows forming a ``star", with energy $4J_2$. As in the Kasteleyn transition, excitations with a finite density of defects are completely suppressed at low temperature. However, as we shall see, there is a temperature-dependent entropic gain associated to these lines due to these excitations, and this entropic gain eventually becomes larger than the entropic cost of a simple DDW. As a consequence, the free energy of a \sC{} line decreases as the temperature is raised up to a critical temperature where it becomes negative. Above this critical temperature a finite density of \sC{} lines develops, driving a phase transition where the antiferromagnetic long-range order of the dense rows is destroyed.

\noindent\textbf{\textit{Topological devil's staircase.}}  
A numerical investigation using a directional version of the Corner Transfer Matrix Renormalization Group (CTMRG) algorithm \cite{sm, nishino_corner_1996,vanhecke_solving_2021,ueda_snapshot_2005} (which can capture highly anisotropic correlations, see~\ref{section:tensor_networks}) confirms that a transition where defect lines condense indeed takes place. 
Note that the density $n_A$ of \sA{} lines is equal to that of crossed hexagons (in the dual bond representation), while the density $n_C$ of \sC{} lines is equal to that of ferromagnetic nearest-neighbor pairs along a dense row. Both can thus be written as local observables, hence easily calculated using CTMRG.\par
Let us look in more details at what happens above the transition. If it was a usual Kasteleyn transition, $n_C$ should grow continuously. However, this is clearly not the case, as shown in Fig.~\ref{fig:2}: $n_C$ jumps to a strictly positive value at the transition $T/J_2 =1.78$. At the same time, $n_A$ drops to a value strictly smaller than 1/3, and most remarkably the ratio $n_C/n_A$ is rigorously equal to 1. The interpretation is quite clear: since the two types of lines must avoid each other, \sC{}-lines must appear between pairs of \sA{}-lines. The configuration realized in the first plateau just above the transition corresponds to exactly one \sC{}-line between two consecutive \sA{}-lines, as confirmed by snapshots 
sampled using CTMRG \cite{sm}, a consequence of the effective repulsion between adjacent \sC{} strings since excitations along consecutive \sC{} lines (not separated by an \sA{}) constrain each other. Note that the densities themselves are slightly temperature dependent.


\begin{figure*}
    \centering
    \includegraphics[width=0.93\textwidth]{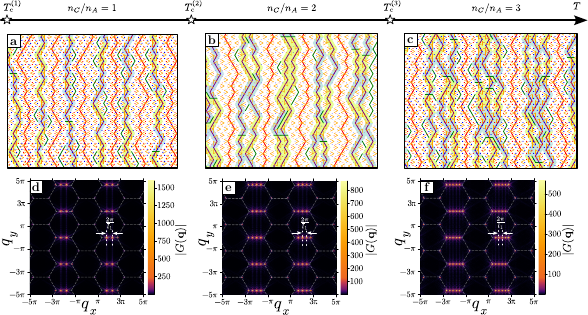}
    \caption{Topological plateaus are shown for increasing temperatures.(a-c): snapshots sampled from the CTM environment ($\chi=80)$ for (a) $ p = n_C/n_A = 1$ ($T=3 J_2$) (b) $p= 2$, $T=3.4 J_2$ (c) $ p = 3$, $T=3.8 J_2$ plateaus, with \sC{}-lines and green decorations. The number $p$ of \sC{}-lines between \sA{} strings is easily evaluated as the number of vertical orange bonds crossing a dense row of spins. Second row (d-f): corresponding magnetic structure factors exhibiting $n_C/n_A+2$ peaks at distances $2\pi/\bar{l}_1$. The structure factors are obtained by averaging over more than 2700 snapshots of size $40\times 40$ and clearly highlight the anisotropy of correlations, also evident in the real-space plots ~\cite{sm}.}
    \label{fig:3}
\end{figure*}

Now,
the restoration of the $\mathbb{Z}_3$ symmetry requires the formation of a 2D graph where \sC{}-lines can develop simultaneously in all directions, to form domains with different orientations of arrows~\cite{smerald_topological_2016}. This can only occur if the \sA{}-lines have completely disappeared,
hence if the ratio $n_C/n_A$ diverges at the transition. With the assumption, numerically validated (see Fig.~\ref{fig:3} a to c), that the number of \sC{}-lines is the same between all pairs of consecutive \sA{}-lines, this implies that this ratio takes integer values from 0 ($\mathbb{Z}_2$ ordered phase) to infinity (high temperature phase). This is supported by our numerical data, where the phases up to a ratio equal to 4 could be unambiguously resolved. Higher plateaus are beyond the reach of our numerics~\cite{sm}. 

We are left with the remarkable conclusion that there is an infinite sequence of first-order phase transitions before the $\mathbb{Z}_3$ symmetry is restored, at which all physical quantities (energy, densities of domain walls, order parameters) jump. This is reminiscent of the devil's staircase observed in the ANNNI and other models where the wave vector
jumps between commensurate values until the high-temperature phase is reached. The mechanism at play in our case is, however, qualitatively different since the density of strings, hence the average distance between domain walls or the wave-vectors of the intermediate phases, are not constant inside the plateaus (see insets of Fig.\ref{fig:2}c).
This devil's staircase is thus not related to commensurability, but of topological origin: it is the number of \sC{} domain walls between consecutive \sA{} domain that changes in steps of 1 upon increasing the temperature. 

\noindent\textbf{\textit{Magnetic structure factors.}}  In order to further characterize the correlations in each of the phases with quantized $n_C/n_A$, we compute the magnetic structure factor (MSF) given by~\cite{rougemaille_cooperative_2019, skjaervo_advances_2020}
\begin{equation}
    G(\mathbf{q}) = \frac{1}{N}\left|\sum_{i=1}^N \sum_{j=1}^{N}\left<\sigma_i\sigma_j\right>_Te^{i\mathbf{q}\cdot(\mathbf{r}_i-\mathbf{r}_j)}\right|.\label{eq:msf}
\end{equation}
Finite-temperature spin correlations are calculated by averaging over large numbers (around 3000) of system snapshots sampled from CTMRG. The MSF calculated for the first three plateaus, shown in Fig.~\ref{fig:3} (d-f), reveals a striking pattern: in the phase with $n_C/n_A=p\in \mathbb{N}$, the diffraction pattern is dominated by a series of equally spaced peaks on the edges of the Brillouin zones. The inter-peak reciprocal distance is equal to $2\pi/\bar{l}_1$, where $\bar{l}_1=1/n_A$ is the average distance between \sA{} lines. If $p$ is odd, there is a peak at $2\pi$ because the antiferromagnetic order around consecutive \sA{} lines is the same, while if $p$ is even, there is no peak at $2\pi$ but a pair of peaks at $2\pi -\pi/\bar{l}_1$ and $2\pi +\pi/\bar{l}_1$ because the antiferromagnetic order is opposite around consecutive \sA{} lines. Finally, all peaks on the edge of the Brillouin zone have significant weight except that at one side of the edge, resulting into $p+2$ peaks for not too large values of $p$, for which $n_A\simeq 1/(p+3)$. All these results can be understood by a detailed analysis of the structure factor of the dense chains and of the $J_3$ sublattice of sparse rows~\cite{sm}.
These MSFs with splitted peaks are reminiscent of structure factors in other aperiodic models, such as incommensurate phases or quasicrystals~\cite{vansmaalen_incommensurate_1995}.

\noindent\textbf{\textit{Discussion.}} To conclude, let us briefly discuss the implications of the present results. The core ingredient of the topological devil's staircase reported here is the existence of two types of line defects in a constrained model, one that has a finite density in the ground state, the other one that can only develop at a finite temperature by an entropic mechanism, as in the Kasteleyn transition. With the additional assumption that these lines repel each other, a standard one for entropically induced defect lines, the conclusion that the transition takes the form of a devil's staircase, in which the number of additional lines between consecutive ground state ones takes all integer values, is quite generic. Indeed, one can show that a statistical physics model of line defects based on these simple assumptions is able to reproduce the main features of the constrained Ising model studied here. Similarly to spin ice~\cite{jaubert_three-dimensional_2008, powell_classical_2008}, a simple quantum-one-dimensional analogue of the present mechanism would be to interpret one of those lines (C-strings) as bosonic worldines. In this analogy, the other type of particles (A) induces a finite-size confinement for the first ones, leading to the observed staircase of jumps as they are slowly driven apart by the increasing chemical potential.
The same physics is also present in a simple extension of the 1D Hubbard model with two types of fermions, the role of the temperature being played as usual by the chemical potential. 
All these related models are reminiscent of the wide variety of origins of the Kasteleyn mechanism, related to many experimental realizations~\cite{nagle_dimer_1989}, and may open paths to observe this unusual series of transitions in magnetic oxides, artificial spin systems, or Rydberg atom arrays.
Finally, the present results also have very interesting consequences for the more realistic model with finite $J_1$ and $J_3$. As long as $J_2$ is not too large, the first transitions turn into very sharp crossovers, with in particular a remarkable series of sharp peaks in the low temperature specific heat. All these extensions will soon be reported elsewhere.
In a broader context, considering recent progress in exploring and understanding constrained models~\cite{placke_ising_2024, zhang_bionic_2024}, we may expect to find other exotic transitions related to the Kasteleyn mechanism~\cite{szabo_perfectly_2024}. This exotic ``classical'' physics may also form a fertile ground for quantum models~\cite{zhang_bionic_2024,powell_quantum_2022, henley_classical_2004,stahl_towards_2025}.\\
\noindent\textbf{\textit{Data availability.}} The data supporting the findings of this
article will be made openly available on 21 Jan. 2027 \cite{dataAvail}.

\noindent\textbf{\textit{Acknowledgements.}} We acknowledge useful discussions with T. Giamarchi, P. Holdsworth, C. Nisoli, B. Vanhecke, L. Vanderstraeten, and F. Verstraete, and the hospitality of the Centro de Ciencias de Benasque Pedro Pascual. Numerical computations have been performed using the facilities of the Scientific IT and Application Support Center of EPFL (SCITAS). 
This work is supported by the Swiss National Science Foundation Grant No. 182179.

\setcounter{section}{0}
\setcounter{secnumdepth}{3}
\setcounter{figure}{0}
\setcounter{equation}{0}
\renewcommand\thesection{EM\arabic{section}}
\renewcommand\thefigure{EM\arabic{figure}}
\renewcommand\theequation{EM\arabic{equation}}
\begin{center}
    \bfseries\large End matter
\end{center}
\section{Local constraints and mapping to strings}
\label{sec:mapping}
\begin{figure}
    \centering
    \includegraphics[width=\linewidth]{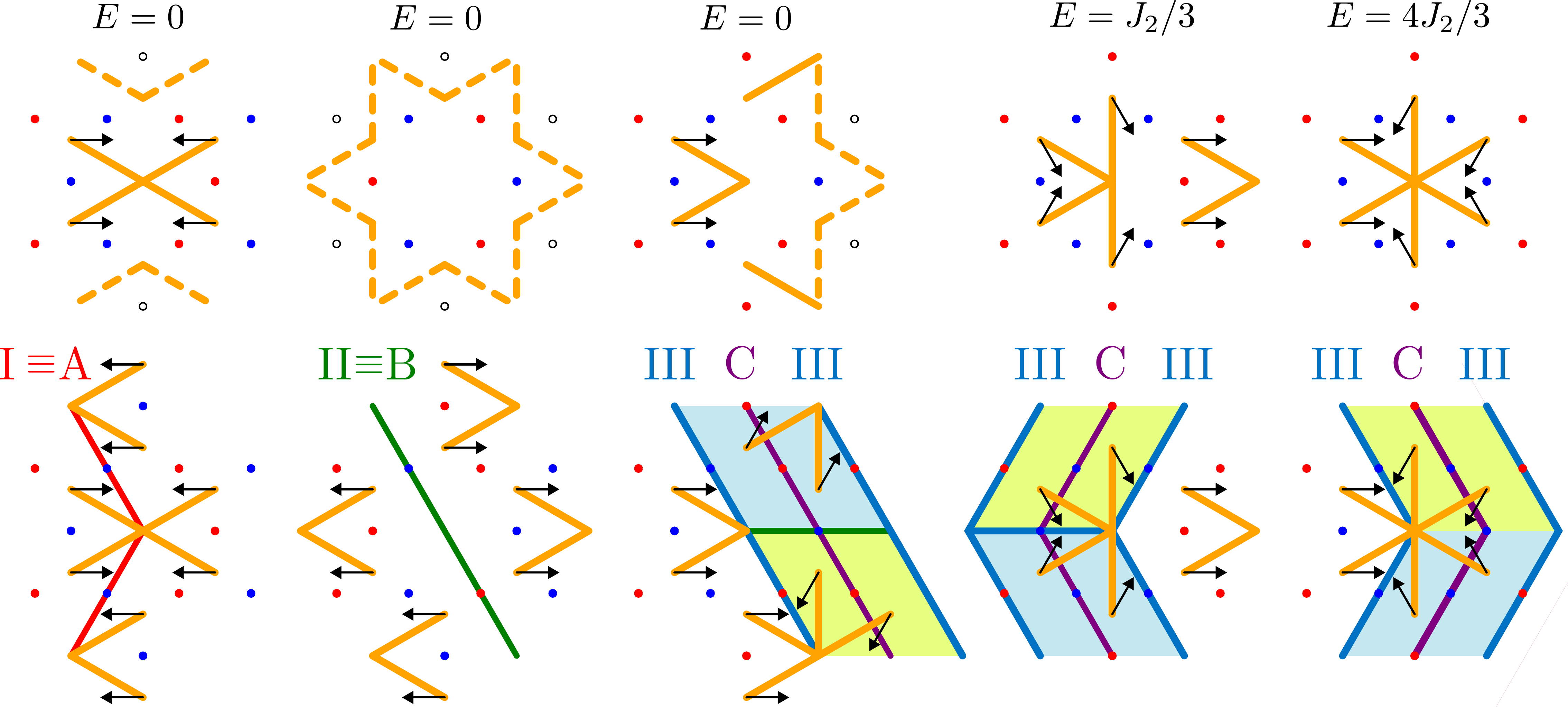}
    \caption{Local configurations of the kagome Ising antiferromagnet in the constrained limit of infinite $J_1$ and $J_3$. Top row: spins on white sites can be flipped with no cost, in which case the two possible alternatives for the dual bonds are dashed. The honeycomb lattice six-state clock-model mapping (black arrows) is defined by attributing to each triangle the direction of the arrow touching it. For the star (rightmost panel) there would be two possible choices; we systematically chose this one by convention. Bottom row: examples of mapping to DWs. Close to hexagons with K's or stars, different spin configurations map onto the same string configuration, contributing to the entropic gain of \sC{} lines.}
    \label{fig:gs_tiles} 
\end{figure}

We show how to map the configurations of the constrained kagome Ising antiferromagnet with infinite $J_1$ and $J_3$ interactions onto the trajectories of three kinds of domain walls. 
The constraints imposed by the $J_1,J_3\rightarrow\infty$ limit forbid ferromagnetic $J_1$ or $J_3$ triangles. As a consequence, spin configurations in each of the three triangular sublattices of $J_3$ bonds must be ground-states of the nearest-neighbor, triangular lattice Ising antiferromagnet. Intuitively, this leads to the extended defects description of the configurations in the constrained manifold: constraint-allowed configurations in each $J_3$ sublattice can be divided into topological sectors that can only be connected by non-local updates \cite{smerald_topological_2016,nourhani_communicating_2018,yokoi_dimer_1986}. A more detailed explanation of the nature of these extended defects, and a proof that they must be system-spanning, is given in the SI \cite{sm}. Here, we focus on the definition of mapping from the spin configuration to the strings. The minimization of $J_1$ and $J_3$ interactions imposes local constraints on the kagome star clusters, allowing the five classes of local configurations listed in Fig.~\ref{fig:gs_tiles}.
As stated in the main text, each nearest-neighbour triangle is crossed by exactly one dual lattice bond, meeting at least one other bond at the center of a hexagon to form an arrow. Arrows form domains with identical orientations, enabling the identification of strings as domain walls. In practice, we map the kagome Ising configurations onto six-state clock model configurations by attributing to each triangle of the kagome lattice the direction of the arrow piercing it (see Fig.~\ref{fig:gs_tiles} for the required conventions). This mapping is not surjective: the ground-state rules of the strings phase impose constraints on the allowed clock model configurations. \par 
The clock-model configuration splits the system into ferromagnetically aligned domains separated by DWs. We distinguish three different classes of DWs:
\begin{itemize}
    \item \textbf{I}: zero-energy DWs between domains with converging parallel directions, in red. 
    \item \textbf{II}: DWs between domains with diverging parallel directions, in green.
    \item \textbf{III}: Walls between domains with directions that differ by 60 or 120 degrees, in blue.
\end{itemize}
The three types of domain walls are drawn in illustrative configurations in the bottom of Fig.~\ref{fig:gs_tiles}. Domain walls of types I and II correspond directly to the \sA{} and \sB{} strings whose trajectories describe the partially ordered family of states mentioned in the main text. Type III domain walls can be combined to form double domain walls (DDW's), which surround columns of oblique arrows. We call each such column a \sC{}-string, drawn in purple.

\section{Tensor networks}
\label{section:tensor_networks}
\begin{figure}
    \centering
    \includegraphics[width=0.8\linewidth]{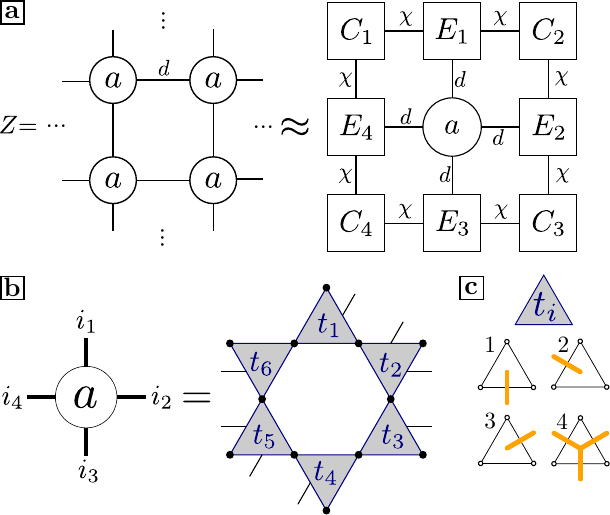}
    \caption{Tensor network. (a) CTRMG idea: the environment surrounding a cluster is approximated by corner ($C_i$) and edge ($E_i$) tensors, which are truncated to a cutoff dimension $\chi$. We use single-site directional CTMRG \cite{orus_simulation_2009} with the isometries introduced in Ref.~\cite{corboz_2014_isometries}. (b) Each tensor in the network contains the Boltzmann weight of a kagome star \cite{vanhecke_solving_2021}. Contraction of the legs of the tensors impose the matching of spin configurations on overlapping sites. (c) If configurations related by a global spin-flip symmetry are identified, there are 4 different states for each triangle. We represent them by drawing a dual lattice bond between adjacent spins pointing in the same direction. The last one is not allowed in the constrained limit ($J_1,J_3\rightarrow\infty$).}
    \label{fig:ctmrg}
\end{figure}

The partition function of a classical lattice model can be calculated as the contraction of a tensor network (TN), an idea directly related to transfer matrix studies of 2D Ising magnets \cite{kramers_statistics_1941,kramers_statistics_1941-1,nishino_review_2022}. 
The exact contraction of a two-dimensional TN is exponentially hard. We use an approximate contraction algorithm, the Corner Transfer Matrix Renormalization Group (CTMRG) \cite{baxter_dimers_1968,baxter_corner_1981,nishino_corner_1996}. In CTMRG, the infinite environment surrounding a site is approximated by the product of corner ($C_i$) and edge ($E_i$) transfer matrices, truncated to a finite cutoff bond dimension $\chi$, as shown in Fig. \ref{fig:ctmrg}a. The directional CTMRG algorithm \cite{orus_simulation_2009} we use does not rely on any assumption of rotational or parity symmetries and has no issue capturing highly anisotropic correlations \cite{Nyckees}.
In the course of CTMRG, tensors are iteratively added to the network and renormalized into the corner and edge tensors until a fixed point is reached. 
Truncation to $\chi$ limits long-range correlations, inducing a finite correlation length even in critical systems.\par

In frustrated models, the impossibility of satisfying every interaction may lead to numerical instabilities in CTMRG at low temperatures and prevent convergence~\cite{vanhecke_solving_2021, song_general_2023}. A solution is to choose a different TN formulation, such that the constraints are enforced within a single tensor. To construct this local tensor, we follow the approach of Ref.~\cite{vanhecke_solving_2021} where the Boltzmann weight is factorized over overlapping clusters, with the contraction of the indices in the network imposing the matching conditions. Here, our clusters are kagome stars. Each pair of corner-sharing triangles is shared between two clusters. This would normally give rise to a triangular TN, but, thanks to the redundancy of the matching conditions, two out of three directions suffice, giving rise to the square TN illustrated in Fig.~\ref{fig:ctmrg}b. \par
We further take advantage of symmetry to divide the bond dimension by two: since configurations related by a global spin-flip have the same energy, it is sufficient to keep track of the position of the ferromagnetic bonds (dual bond configurations). Since there are four possible dual bond configurations per triangle (Fig.~\ref{fig:ctmrg}c), the TN has bond dimension 16.

\section{Computing order parameters for rotational and spin-flip symmetries}
\label{sec:oderparameter}

As mentioned in the main text, the ground states of the strings phase are dominated by a partially ordered family with antiferromagnetic long-range order on the dense rows and quasi-long-range order on the sparse rows. Thus, the strings phase breaks the $\mathbb{Z}_3$ rotational symmetry and the $\mathbb{Z}_2$ spin-flip symmetry at low temperatures. 
We numerically detect the restoration of these symmetries as follows. First, the initial tensors for the environment are chosen such that the $\mathbb{Z}_3$ symmetry is slightly broken by favoring a certain orientation of the crosses (this is the equivalent of slightly favoring up spins in the initial condition for a TN simulation of the 2D square lattice Ising model: it only selects a finite magnetization if there is spontaneous symmetry breaking.).

This allows us to compute the $\mathbb{Z}_3$ order parameter using a local tensor which detects whether a particular dual bond is favored:
\begin{equation}
    \begin{aligned}
        \Psi_{\mathbb{Z}_3}&=\left|\left<\sigma_1(\mathbf{0})\sigma_2(\mathbf{0}) +\sigma_2(\mathbf{0})\sigma_3(\mathbf{0})e^{2i\pi/3}\right.\right.\\
         &+\left.\left.{\sigma_3(\mathbf{0}) \sigma_1(\mathbf{0}) e^{-2i\pi/3}}\right>_T\right|,\label{eq:z3_op}
    \end{aligned}
\end{equation}
where $\{\sigma_1(\mathbf{R}),\sigma_2(\mathbf{R}),\sigma_3(\mathbf{R})\}$ are the three spins at the vertices of a triangle centered around $\mathbf{R}$. This $\mathbb{Z}_3$ order parameter is related to the density of of \sC{} strings by
$\Psi_{\mathbb{Z}_3}=1/2-3n_C/4$,
which results from reflection symmetry and the constraints imposed by
infinite $J_1$.

The order parameter for the $\mathbb{Z}_2$ antiferromagnetic order cannot be computed using only a local tensor since the spin-flip symmetry is enforced at the level of the TN (see Sec.~\ref{section:tensor_networks}). However, with the fixed boundary conditions, we can compute it as the limit of the spin-spin correlation function along a dense row
\begin{align}
    \Psi_{\mathbb{Z}_2}&=\lim_{r\rightarrow\infty} \sqrt{|\left<\sigma_{1}(\mathbf{0})\sigma_{1}(r\mathbf{e}_x)\right>_T|} \label{eq:z2_op}.
\end{align}

\begin{figure}
    \centering
    \includegraphics[width=0.8\linewidth]{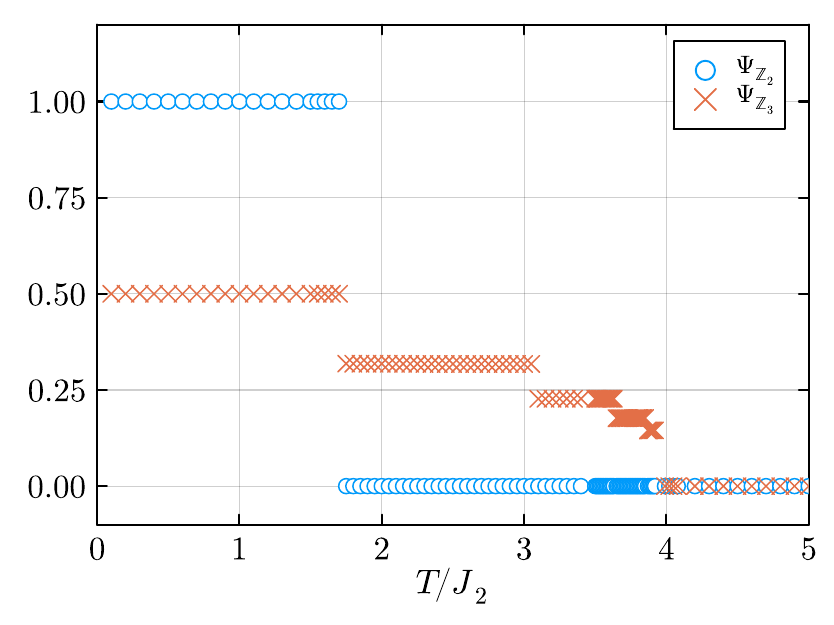}
    \caption{CTMRG results ($\chi=100$) for the temperature dependence of the order parameters for the $\mathbb{Z}_2$ and $\mathbb{Z}_3$ broken symmetries, with infinite $J_1$ and $J_3$. Only temperatures where the results are converged in bond dimension are shown.}
    \label{fig:order_parameters}
\end{figure}

The numerical calculation of both order parameters are shown in Fig. \ref{fig:order_parameters}. $\mathbb{Z}_2$ is restored at the first phase transition ($T_c^{(1)}=1.78 J_2$), while the $\mathbb{Z}_3$ symmetry is only restored at the accumulation point ($T_c^{(\infty)}=3.99 J_2$), which means that the phases between these temperatures are nematic.

\bibliography{references}

\setcounter{section}{0}
\setcounter{secnumdepth}{3}
\setcounter{figure}{0}
\setcounter{equation}{0}
\renewcommand\thesection{S\arabic{section}}
\renewcommand\thefigure{S\arabic{figure}}
\renewcommand\theequation{S\arabic{equation}}

\newpage
\onecolumngrid
\begin{center}
    \bfseries\Large Supplementary Information
\end{center}

\section{Snapshot sampling using CTMRG}
An often overlooked aspect of CTMRG is its ability to efficiently sample large snapshots of systems according to the equilibrium Gibbs distribution, as demonstrated by Ueda \textit{et al.} in 2005 \cite{ueda_snapshot_2005} for the case of the square-lattice Ising model. The key idea is to sequentially sample spins in a rectangular region. The first spin is sampled at random from the distribution given by the environment. The conditional probability distribution for the next spin given the already sampled ones can be computed as a TN contraction whose cost grows linearly with the snapshot size. The sampling of system snapshots using CTMRG may also have applications in hybrid Tensor-Network Monte-Carlo schemes \cite{frias-perez_collective_2023,chenTensorNetworkMonte2025}. \par
\begin{figure}
    \centering
    \includegraphics[width=\linewidth]{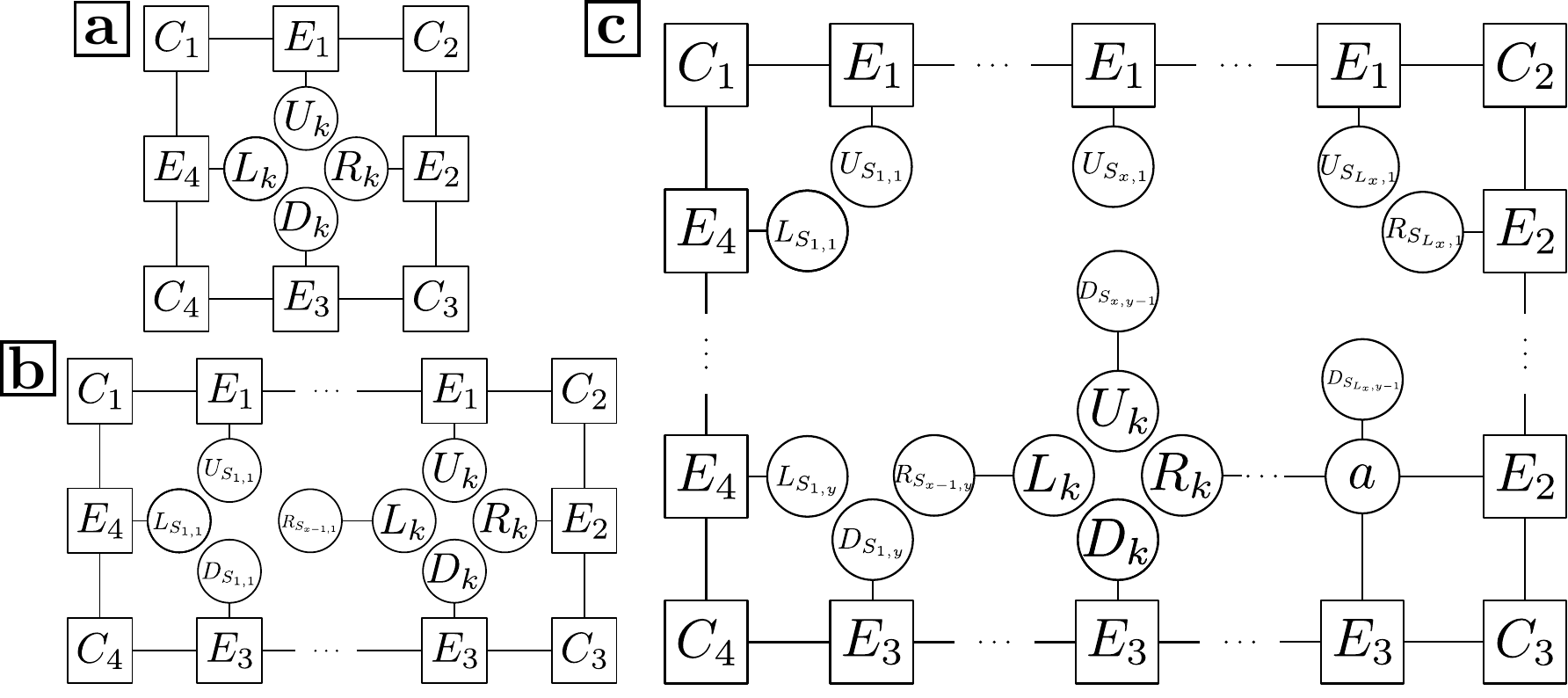}
    \caption{Illustration of the algorithm for snapshot sampling with CTMRG. The conditional probabilities distribution for a cluster in (a) the first hexagon, (b) the first row, and (c) any other row, can be calculated by performing a contraction. In order for the probability densities to be properly normalized they must be divided by the sum over the $2^{11}$ different cluster configurations.}
    \label{fig:snapshot_sampling}
\end{figure}

Here, we briefly outline the method, adapted to the setting of the kagome lattice Ising model. We first note that the probability density of a cluster configuration $k$ can be computed using the local tensor
\begin{equation}
    \rho(k)_{i_1,i_2,i_3,i_4} = \delta_{i_1,i_1(k)} \delta_{i_2,i_2(k)} \delta_{i_3,i_3(k)} \delta_{i_4,i_4(k)} e^{-E_k/T}
\end{equation}
which can be factorized as the tensor product of four vectors as illustrated in Fig.~\ref{fig:snapshot_sampling}a
\begin{equation}
   \rho(k)_{i_1,i_2,i_3,i_4}  = (U(k)_{i_1} R(k)_{i_2} D(k)_{i_3}  L(k)_{i_4})\label{eq:prob_factorization}.
\end{equation}
In order to sample a snapshot in a rectangular region of $L_x\times L_y$ clusters, the following procedure is used:
\begin{enumerate}
    \item The probability distribution for a single cluster configuration, $P(k)$, is calculated by contracting the tensors $\rho(k)$ with the converged CTMRG environment and normalizing appropriately.
    \item The first cluster configuration, $S_{1,1}$, is sampled from the distribution $P(k)$. We set the position $(x,y)=(2,1)$.
    \item The conditional probability distribution, $P(S_{x,y}|S_{1,1},\dots,S_{x-1,y})$, of the cluster configuration at position $(x,y)$ given all the previously sampled configurations is calculated by TN contraction, as prescribed in Fig. \ref{fig:snapshot_sampling}(b,c).
    \item The cluster configuration $S_{x,y}$ is sampled from the conditional probability distribution $P(S_{x,y}|S_{1,1},\dots,S_{x-1,y})$. If $(x,y)\neq (L_x,L_y)$, we change it to the next cluster following top to down and left to right order, and return to step 3.
\end{enumerate}
Following this procedure, it is possible to sample snapshots with a computational cost which scales as $O(\chi^2L_xL_y)$. The possibility of efficiently sampling system snapshots also provides a practical alternative for numerically calculating the MSF, instead of using channel environments~\cite{Vanderstraeten_gradient_2016}.

\section{Real-space correlations}\label{section:real_space_correlations}
\begin{figure}
    \centering
    \includegraphics[width=\linewidth]{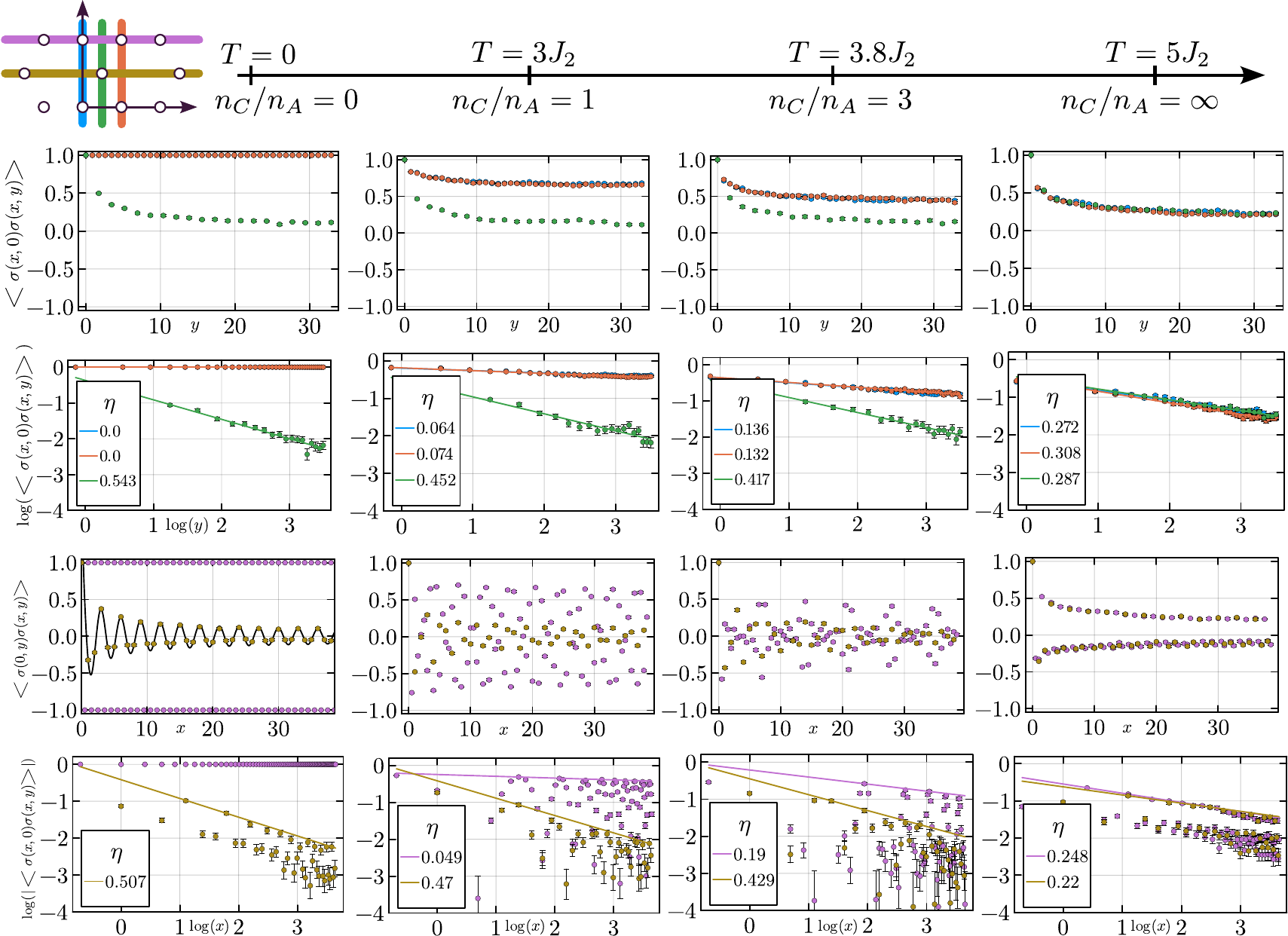}
    \caption{Real-space correlations between spins on different curves of the Kagome lattice at four temperatures, corresponding to phases with different values of $n_C/n_A$. Each color in the correlation plots corresponds to a row or column of the Kagome lattice, as prescribed by the illustration on the top-left. 1st row: Correlation function in the vertical direction. 2nd row: Correlation function in the vertical direction plotted in a log-log scale. The critical exponent $\eta$ is determined from a linear fit. 3rd row: Correlation function in the horizontal direction. On the leftmost panel, the exact spin-spin correlations of the Triangular Ising antiferromagnet  plotted in black~\cite{stephenson_isingmodel_1970}. 4th row: Correlation functions plotted in log-log scale. Due to incommensurability, the critical exponent $\eta$ is determined from the slope of the convex hull of the log-log plots.  Correlations were calculated by averaging over more than 2700 snapshots of size $40\times40$. Error bars indicate statistical uncertainties (whenever error bars are not visible, they are smaller than the data-point markers).}
    \label{fig:correlations}
\end{figure}

As briefly mentioned in the main text, the ability to sample system snapshots permits the calculation of correlations between any two spins. In Fig. \ref{fig:correlations}, the correlations between spins in different rows and columns of the Kagome lattice are shown at different temperatures. The highly anisotropic nature of spin-spin correlations is evident in the fact that vertically separated correlations are positive and monotonously decreasing, while horizontal correlations are modulated by commensurate or incommensurate wave-vectors (depending on temperature). Looking at the plots in a log-log scale, correlations can be noticed to decay algebraically with distance at all temperatures. \par 
The algebraic correlations were further characterized by the $\eta$ critical exponent, defined by the asymptotic expression
\begin{equation}
    G(r)\sim f_{\text{osc.}}(r)r^{-\eta}\Rightarrow \log G(r) \sim \log \left\{f_{\text{osc.}}(r)\right\} -\eta\log r\label{eq:eta_exponent},
\end{equation}
where $1\leq f_{\text{osc.}}(r)\leq 1$ is a an oscillating function, zero on average, and which does not decay with distance. In the non-oscillating case ($f_{\text{osc.}}(r)=1$) $\eta$ is determined through a linear fit to the log-log correlation curve. In the oscillating case, due to the challenges resulting from incommensurability, we estimate $\eta$ from the slope of the upper convex hull of the absolute value of correlations in log-log scale: at large distances, the upper convex hull follows the amplitude of oscillations, equal to evaluating $G(r)$ in a sequence of points where $|f_{\text{osc.}}(r)|\approx 1$. The determination of the $\eta$ exponent reveals that  correlations between spins on dense rows decay faster at higher temperatures, while the correlations between spins on sparse rows decay faster at lower temperatures. \par

\section{Dependence of results on bond dimension}
The precision of the Corner Transfer Matrix Renormalization Group (CTMRG) is controlled by the cutoff bond dimension $\chi$. $\chi$ limits the range of correlations which can be captured, and therefore induces a finite correlation length even in critical systems. This finite correlation length may in turn be interpreted as defining an effective system size. As in the case of finite-size scaling, scaling hypothesis for the bond dimension may therefore be used in order to study the properties of critical systems \cite{BramMPSscaling,rams_precise_2018}.\par
\begin{figure}
    \centering
    \includegraphics[width=0.7\linewidth]{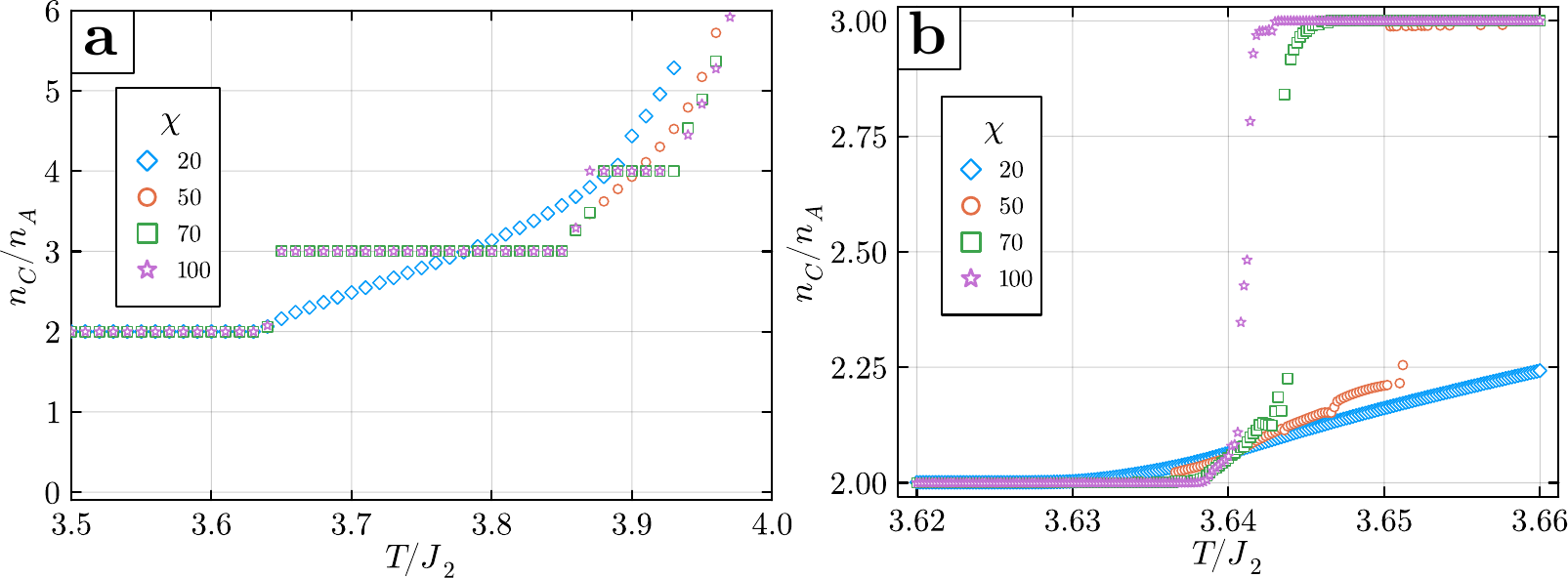}
    \caption{Ratio of string densities $n_C/n_A$ as a function of temperature, calculated using CTMRG with three different bond dimensions. The converged environment with smaller $\chi$ was used to initialize the calculation with the next value of $\chi$ in the sequence $\chi\in \{10,20,\dots,100\}$. (a) Plateaus with larger integer ratios become visible only at larger bond dimensions. (b) The smooth crossover between plateaus gets narrower as bond dimension is increased.}
    \label{fig:conv_bond_dim}
\end{figure}

As discussed in section \ref{section:real_space_correlations}, our numerical results show that the constrained limit of the Kagome Ising antiferromagnet is critical at all temperatures. As a consequence of criticality, observables computed with CTMRG on the constrained model display strong bond dimension dependence. If the ratio of densities $n_C/n_A$ is plotted with different values of $\chi$ (Fig. \ref{fig:conv_bond_dim}a), we can observe that the plateau with $n_C/n_A=4$ is present only if $\chi\geq70$, while plateaus with smaller integer ratios are present even at lower bond dimensions. To understand why, we recall that spin correlations in the quantized phases are \emph{modulated} by a characteristic length equal to the average distance between \sA{} strings: $\bar{l}_1=1/n_A\approx n_C/n_A+3$. This distance increases with the value of the integer ratio, which means that the higher-temperature plateaus can only be accurately described by CTMRG at larger bond dimensions. 
Note that the numerical correlation length is much larger than $\overline{l}_1$.
\par
Finally, we also discuss the presence of points on the boundary between the phases, for example at $T/J_2=3.64$, which seem to violate the quantization of $n_C/n_A$ to integer values. A finer grained calculation around this temperature (Fig. \ref{fig:conv_bond_dim}) reveals that these points belong to the crossover between adjacent plateau phases. The width of the crossover is narrowed as $\chi$ is increased, exactly as in a finite system near a first-order phase transition. The numerical results support the hypothesis that in the thermodynamic limit ($\chi\rightarrow\infty$) $n_C/n_A$ is exactly quantized to integer values.

\section{Approximate calculations of the magnetic structure factor}
\begin{figure}
    \centering
    \includegraphics[width=0.7\linewidth]{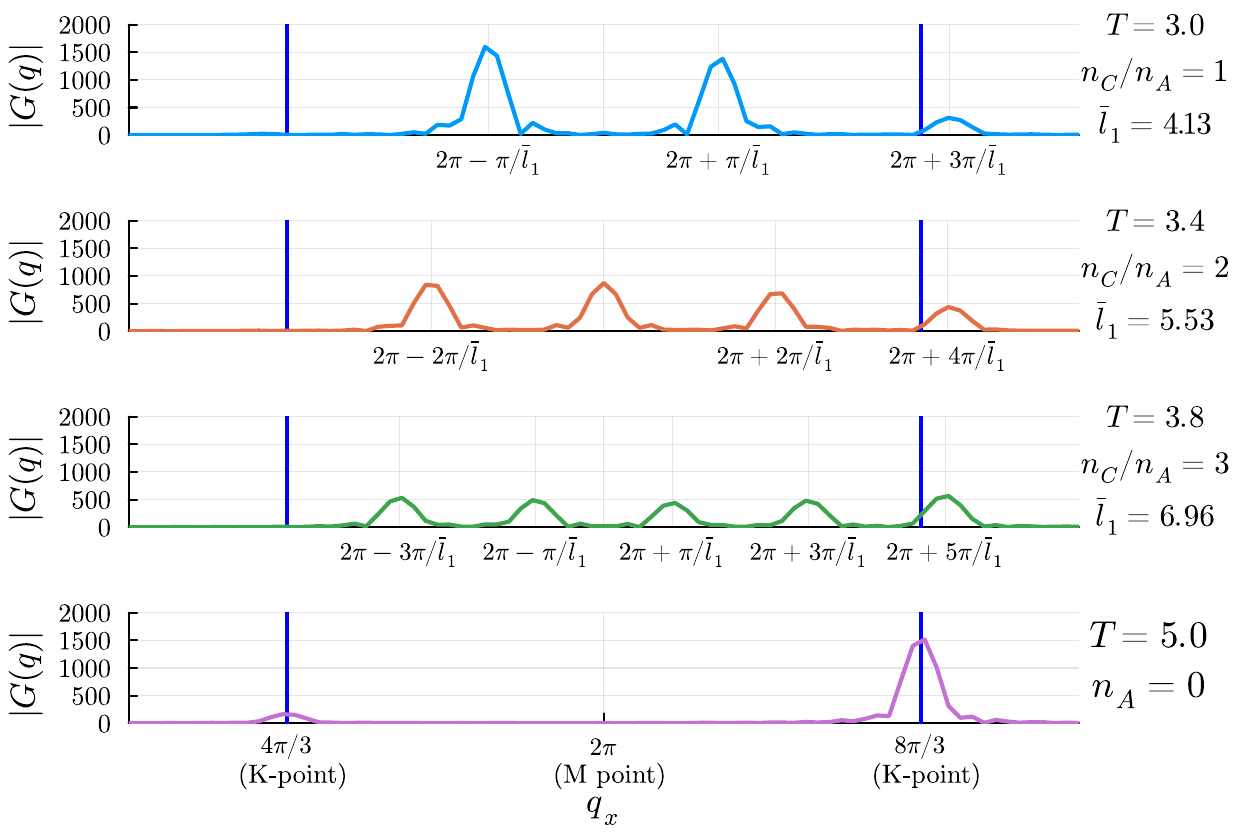}
    \caption{Plots of the MSF of the $J_1,J_3\rightarrow\infty$ constrained model along the $(h,0,0)$ direction. }
    \label{fig:msf_cuts}
\end{figure}

As discussed in the main text, the MSF of the intermediate phases of the constrained kagome Ising antiferromagnet with $J_1,J_3\rightarrow\infty$ satisfies a remarkable property: in the phase with $n_C/n_A=p$, $p+2$ peaks are visible on the edge of the Brillouin zone. The position and relative intensities of the different peaks is clear in the cut of the MSF plotted along the $\mathbf{q}=(q_x,0,0)$ line in Fig. \ref{fig:msf_cuts}.\par
Here, we perform a simplified calculation to understand this property. First, we write the structure factor as a sum over sublattices that will allow us to discuss separately the contributions of the dense and sparse rows:
\begin{align}
    G(\mathbf{q})&=\sum_{a=1}^3\sum_{b=1}^3 e^{i\mathbf{q}(\boldsymbol{\tau}_a-\boldsymbol{\tau}_b)}G_{a,b}(\mathbf{q})\\
    G_{a,b}(\mathbf{q})&=\lim_{L\rightarrow\infty}\frac{1}{3L^2}\sum_{(x_0,y_0)\in(\mathbb{Z}_L)^2}\sum_{(x,y)\in(\mathbb{Z}_L)^2}e^{i((x-x_0)\mathbf{a}_1+(y-y_0)\mathbf{a}_2)\cdot\mathbf{q}}\left<\sigma_a(x_0,y_0)\sigma_b(x,y)\right>_T,
\end{align}
where the sub-lattice index $a\in\{1,2,3\}$ in $\sigma_a(x,y)$ refers to a spin on a sparse row if $a=3$ and to a spin on a dense row otherwise. The vectors $\boldsymbol{\tau}_1 = 0$, $\boldsymbol{\tau}_2 = \mathbf{a}_1/2$, and $\boldsymbol{\tau}_3 = (\mathbf{a}_1 + \mathbf{a}_2)/2$ specify the positions of the three sites within each unit cell of the kagome lattice.

\par
Let us first focus on the contribution of the spins on dense rows to the MSF. In the constrained model, the configuration of spins on dense rows depends solely on the position of the \sC{} lines. We work under the approximation that \sC{} lines are perfectly vertical, setting the distance between them to the incommensurate, average distance. Then, the spin-spin correlations are independent of the vertical coordinate. We get

\begin{equation}
    \begin{aligned}
        \sum_{(a,b)\in\{1,2\}^2}e^{i\mathbf{q}\cdot(\boldsymbol{\tau}_a-\boldsymbol{\tau}_b)}G_{a,b}(\mathbf{q})&=\sum_{(x,y)\in\mathbb{Z}^2}\sum_{(a,b)\in\{1,2\}^2}e^{iq_x\left(x+\frac{a-b}{2}\right)}e^{iq_y y\sqrt{3}/2}G_{\text{row}}\left(x+\frac{a-b}{2}\right),\label{eq:msf_dense_rows}
    \end{aligned}
\end{equation}
where $G_{\text{row}}$ is the spin-spin correlation function within a dense row. Since $y$ is not coupled to $a$ and $b$, this expression factorizes into a sum over $y$ and a sum over $x$. 

\begin{figure}
    \centering
    \includegraphics[width=\linewidth]{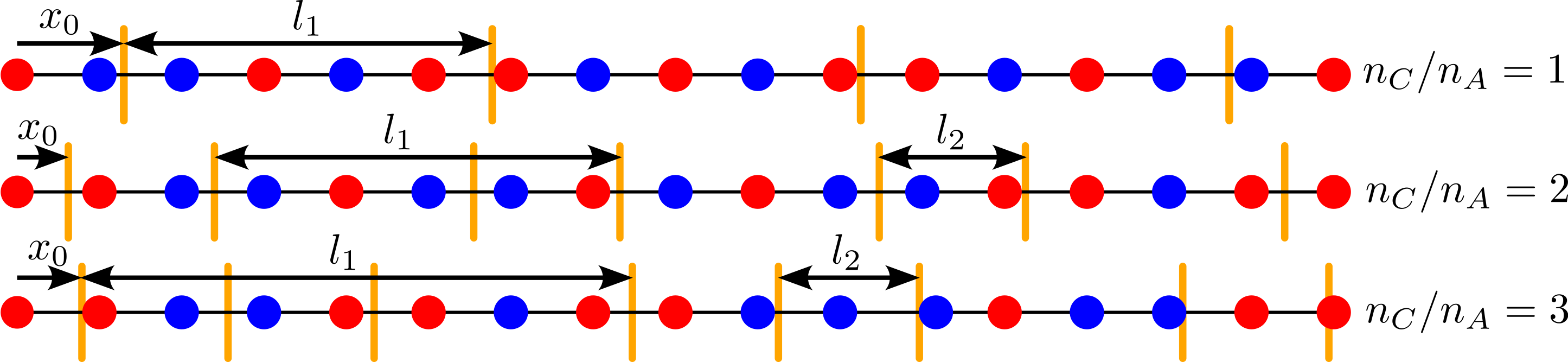}
    \caption{Illustration of the model of uniformly spaced domain walls leading up to the correlation function for three different values of $n_C/n_A=p$. $\bar{l}_1$ is the overall periodicity of the Domain Wall arrangement, while $l_2$ is the distance between nearest-neighbor domain walls. The spin-spin correlation function can be calculated by averaging over the position of the first domain wall - $x_0$.}
    \label{fig:msf_model}
\end{figure}

In order to evaluate the sum over $y$, we use the formula for the Fourier transform of the Dirac comb \cite{bracewellFourierTransformIts1986}
\begin{equation}
    \sum_{n\in\mathbb{Z}}e^{in\omega}= 2\pi \sum_{m\in \mathbb{Z}}\delta(\omega-2\pi m),\label{eq:dirac_comb}
\end{equation}
while in the sum over $x$, the sum over $a$ and $b$ can be performed explicitly, leading to
\begin{equation}
    \sum_{(a,b)\in\{1,2\}^2}e^{i\mathbf{q}\cdot(\boldsymbol{\tau}_a-\boldsymbol{\tau}_b)}G_{a,b}(\mathbf{q})=\left(2\pi\sum_{m_2\in\mathbb{Z}}\delta\left(q_y-\frac{4\pi m_2}{\sqrt{3}}\right)\right)\left(2\sum_{u\in\mathbb{Z}}e^{iq_xu/2}G_{\text{row}}(u/2)\right).
\end{equation}
Let us start by treating the case of the first plateau, where there is a single \sC{} line  between pairs of \sA{} lines. The average distance between domain walls $\bar{l}_1$ is incommensurate. In this plateau, it varies around the value
\begin{equation}
    \bar{l}_1=\frac{1}{n_A}\approx1/0.2425=4.124.
\end{equation}
Neglecting fluctuations, we approximate the domain walls to be evenly spaced by $\bar{l}_1$, as shown in Fig. \ref{fig:msf_model}. The spin-spin correlation function is therefore
\begin{equation}
    G^{p=1}_{\text{row}}(x)=e^{2i\pi x}\int_0^{\bar{l}_1}\frac{dx_0}{\bar{l}_1}\exp\left\{i\pi \left\lfloor\frac{x+x_0}{\bar{l}_1}\right\rfloor\right\},\label{eq:g_1_real}
\end{equation}
where $f(x) =\lfloor x\rfloor$ is the floor function (integer part of a real number, rounded down). In the previous equation, the oscillating factor $e^{2i\pi x}=(-1)^{2x}$ encodes the underlying antiferromagnetic order (note that $x$ can take integer or half-integer values), while the factor $\exp\left\{i\pi \left\lfloor\frac{x+x_0}{\bar{l}_1}\right\rfloor\right\}$ accounts for the stacking faults which appear whenever there is a domain wall between two adjacent spins. $x_0$ is the distance from the reference spin to the first domain wall to its right, and averaging over $x_0$ is equivalent to averaging over the reference spin, translating it along the chain.\par
\indent In order to calculate the Fourier transform of Eq. \eqref{eq:g_1_real}, we use that  the function $\exp\left\{i\pi\left\lfloor \frac{x+x_0}{\bar{l}_1}\right\rfloor\right\}$ has a period of $2\bar{l}_1$ and express it as the Fourier series
\begin{equation}
    \exp\left\{i\pi\left\lfloor \frac{x+x_0}{\bar{l}_1}\right\rfloor\right\}=\sum_{k=1}^{\infty}\frac{4}{\pi(2k-1)}\sin\left(\frac{(2k-1)\pi (x+x_0)}{\bar{l}_1}\right).
\end{equation}
The row correlation function thus becomes
\begin{equation}
\begin{aligned}
    G^{p=1}_{\text{row}}(x)&=e^{2i\pi x}\int_0^{\bar{l}_1}\frac{dx_0}{\bar{l}_1}\left[\sum_{k=1}^{\infty}\frac{4}{\pi(2k-1)}\sin\left(\frac{(2k-1)\pi (x+x_0)}{\bar{l}_1}\right)\right]=\\
    &=e^{2i\pi x}\sum_{k=1}^{\infty}\frac{8}{\pi^2(2k-1)^2}\cos\left(\frac{\pi(2k-1)x}{\bar{l}_1}\right).\label{eq:g_p_1_real_space}
\end{aligned}
\end{equation}
By combining equations \eqref{eq:msf_dense_rows} and \eqref{eq:g_p_1_real_space}, we obtain finally
\begin{equation}
     \begin{aligned}
     \sum_{(a,b)\in\{1,2\}^2}e^{i\mathbf{q}\cdot(\boldsymbol{\tau}_a-\boldsymbol{\tau}_b)}G_{a,b}(\mathbf{q})&=4\pi\left(\sum_{m_2\in\mathbb{Z}}\delta\left(q_y-\frac{4\pi m}{\sqrt{3}}\right)\right)\left(\sum_{k=1}^{\infty}\frac{4}{\pi^2(2k-1)^2}\left[\sum_{u\in\mathbb{Z}}\exp\left\{\frac{i}{2}\left(q_x+2\pi+\frac{\pi(2k-1)}{\bar{l}_1}\right)u\right\}\right.\right.\\
     &\left.\left.+\sum_{u\in\mathbb{Z}}\exp\left\{\frac{i}{2}\left(q_x+2\pi-\frac{\pi(2k-1)}{\bar{l}_1}\right)u\right\}\right]\right)\\
     &=\sum_{(m_1,m_2,k)\in\mathbb{Z}^3}\frac{32}{(2k-1)^2}\delta^2\left(\mathbf{q}-\frac{4\pi m_2}{\sqrt{3}}\mathbf{e}_y+\left[4\pi m_1+2\pi+\frac{\pi(2k-1)}{\bar{l}_1}\right]\mathbf{e}_x\right),
     \end{aligned}
\end{equation}
where we used Eq. \ref{eq:dirac_comb} in order to evaluate the sum over $u$. The model of uniformly spaced domain-walls leads to Bragg peaks at the wave-vectors
\begin{equation}
    \mathbf{Q}^{p=1}_{(m_1,m_2,k)}=\left(4\pi m_1+2\pi+\frac{\pi(2k-1)}{\bar{l}_1},\frac{4\pi m_2}{\sqrt{3}}\right).\label{eq:peak_list_p1}
\end{equation}
If $\bar{l}_1$ is incommensurate with the underlying lattice spacing, the set of peak positions $\mathbf{Q}^{p=1}_{(m_1,m_2,n)}$ is in principle dense in the horizontal lines defined by $q_y=\frac{4\pi }{\sqrt{3}}m_2$, a common situation for one-dimensional incommensurate systems \cite{vansmaalen_incommensurate_1995,janotQuasicrystalsPrimer1992}. However, the prefactor $1/(2k-1)^2$ strongly suppresses peaks with large values of $k$, and peaks in the vicinity of a given peak at a small value of $k$ would correspond to very large values of $k$ shifted back by a large value of $m_1$, and in practice they are invisible. The one-dimensional model of uniformly spaced domain walls thus accurately predicts that peaks will be visible at $\frac{\pi(2k-1)}{\bar{l}_1}$ for small $k$. However, it does not explain why only the three peaks with $0\leq k \leq 2$ can be observed. In part, the vanishing intensity of peaks with large $|k|$ can be attributed to the randomness of the domain wall positions,  also responsible for smoothening the Bragg peaks. This hypothesis was confirmed numerically by calculating the MSF of a one-dimensional model where the distance between adjacent domain walls is a random variable of average $\bar{l}_1$ and variance $(\delta \bar{l}_1)^2$ (data not shown). However, the intensity asymmetry which leads to the peak with $k=-1$ but not the one with $k=2$ being extinguished is not simply a result of fluctuations. As discussed in more detail later, this results from the correlation with spins on sparse rows. \par
The cases of plateaus with larger ratio $n_C/n_A=p\in\mathbb{N}$ can be worked out similarly. There, the arrangement of domain walls can be better modeled by two characteristic distances: the overall periodicity of the pattern is $\bar{l}_1=1/n_A$, while the distance from each domain wall to its closest neighbor is $l_2$. The spin-spin correlations are calculated by averaging over the position of the first domain wall, leading to
\begin{equation}
    G_{\text{row}}^{p}(x) = e^{2i\pi x}\int_{0}^{\bar{l}_1}\frac{dx_0}{\bar{l}_1} \prod_{q=0}^{p-1}\exp\left\{i\pi\left\lfloor\frac{x+pl_2+x_0}{\bar{l}_1}\right\rfloor\right\}\exp\left\{i\pi\left\lfloor\frac{pl_2+x_0}{\bar{l}_1}\right\rfloor\right\}.\label{eq:g_row_gen_p}
\end{equation}
The Fourier transform of the previous function can once again be calculated by expanding the function $\exp\left\{i\pi\left\lfloor \frac{x}{\bar{l}_1}\right\rfloor\right\}$ as a Fourier series, as shown in the case of $n_C/n_A=1$. It has Bragg peaks at the wave-vectors:
\begin{equation}
    \mathbf{Q}^{p}_{(m_1,m_2,k)}=\begin{cases}
        \left(4\pi m_1+2\pi+\frac{\pi(2k-1)}{\bar{l}_1},\frac{4\pi m_2}{\sqrt{3}}\right)\text{, if p is odd}\\
        \left(4\pi m_1+2\pi+\frac{\pi(2k)}{\bar{l}_1},\frac{4\pi m_2}{\sqrt{3}}\right)\text{, if p is even}
    \end{cases}\label{eq:Q_list}.
\end{equation}
Even though the relative intensity of the Bragg peaks also depends on $l_2$, the position of the peaks only depends on $p$ and $l_1$. As in the case of the $n_C/n_A=1$ plateau, the peak positions predicted by the one-dimensional model agree with the MSF of the constrained kagome Ising antiferromagnet, but the one-dimensional model does not explain why only the $p+2$  peaks with $k$ between $k_{\text{min}}(p)=\lfloor -p/2\rfloor+1$ and $k_{\text{max}}(p)=\lfloor p/2\rfloor+2$ can be observed. The vanishing intensity of peaks with large $|k|$ can again be attributed to the randomness of the domain wall positions,  also responsible for the smoothening the Bragg peaks.\par
The vanishing of the peak with $k=\lfloor -p/2\rfloor$ but not the one with $k=\lfloor p/2\rfloor+2$ cannot be understood solely on the basis of a model for the dense rows, as the symmetry of the rectangular sub-lattice of dense rows requires that $\tilde{G}\left(4\pi m_1+2\pi+q_x,\frac{4\pi m_2}{\sqrt{3}}\right)=\tilde{G}\left(4\pi m_1+2\pi-q_x,\frac{4\pi m_2}{\sqrt{3}}\right)$. Therefore, the spins in the sparse sub-lattice must be included in the description. Since those spins are affected by the position of defects along the \sC{} strings, their correlations cannot be efficiently captured by a one-dimensional model. \par
\begin{figure}
    \centering
    \includegraphics[width=0.7\linewidth]{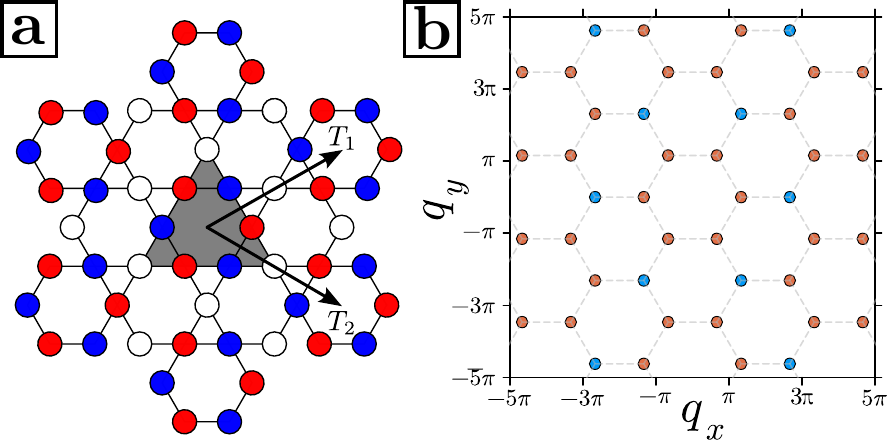}
    \caption{Maximally flippable family of states of the $J_1-J_3$ model. (a) Illustration of the spin configurations in the family of states. White spins may be chosen independently to be up or down. (b) The MSF has \textit{weak}, shown in orange, and \textit{strong}, shown in blue, peaks at the K-points of the Brillouin zones. Strong peaks are nine times stronger than the weak ones.}
    \label{fig:max_flip}
\end{figure}

 
Instead, one may gain insight into the nature of these correlations by realizing that within a \sC{} string of width $p\gg 1$, spin correlations are approximately those of the $J_1-J_3$ kagome Ising antiferromagnet, which is also the $T\rightarrow\infty$ limit of the constrained model. In turn, correlations in the $J_1-J_3$ model are algebraically decaying and modulated by the periodicity of the maximally flippable family of states shown in Fig. \ref{fig:max_flip}, whose Fourier transform leads to peaks of two different intensities at the K-points of the Brillouin zones.
The peaks predicted by the one-dimensional model, listed in Eq. \eqref{eq:Q_list}, are very close to the K-points for $k=\lfloor -p/2\rfloor$ and $k=\lfloor p/2\rfloor+2$. Therefore, analogy with the $J_1-J_3$ model can shine some light on the extinction of the peak with $k=\lfloor -p/2\rfloor$: it is close to the position of one of the \textit{weak} peaks of the $J_1-J_3$ model, while the one with $k=\lfloor p/2\rfloor+2$ is close of the position of a \textit{strong} peak.

\section{Additional ground states of the strings phase}

\begin{figure}
    \centering
    \includegraphics[width=\linewidth]{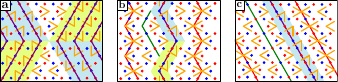}
    \caption{Some configurations that belong to the strings phase ground state manifold cannot be described simply by replacing \sB{} lines with DDWs in the partially ordered family of states described by the trajectories of \sA{} and \sB{} lines. Examples of this include configurations which: a) contain six domains rotated by 60º from each-other and which meet at an empty hexagon; b) contain a single divergence of arrow direction inside a DDW, and possibly a \sB{} line with ends on each side of the divergence; c) contain solely straight domain walls, which allows for at most one \sB{} line and any number of DDWs between a pair of \sA{} lines. The number of configurations in each of these categories grows sub-extensively with system size.}
    \label{fig:exotic_gs}
\end{figure} 
As described in the main text, the partial order in the ground state of the strings phase of the kagome Ising antiferromagnet is due to a family of states corresponding to the trajectories of \sA{} and \sB{} strings. The numerical observation that the density of \sC{} lines is zero in the ground state implies that the partially ordered family occupies a fraction of configurations which approaches one in the thermodynamic limit. This also explains why the number of states in this family is consistent with the residual entropy of the strings phase.

\par
As mentioned in the main text, another important family of ground states can be described by replacing one or more \sB{} lines with Double Domain Walls (DDWs). Even though the size of this second family also grows extensively with system size, the thickness of the DDW's means that they account for a vanishing fraction of states in the thermodynamic limit. If all ground states were in one of these families, this would constitute a proof for the numerical observation that only the states in the partially ordered phase have nonzero probability.\par
However, the completion of the proof is prevented by the existence of states which do not belong to either family. Three examples of such states are shown in Fig. \ref{fig:exotic_gs}. They are representative of classes of states whose number, although infinite in the thermodynamic limit, grows only subextensively with the size of the system. Due to the numerical evidence obtained from CTMRG, we strongly believe that there is only a subextensive number of ground states outside the two main families described in the main text. Still, since the complexity of the ground-state rules prevents extensive enumeration, this statement falls short of being an exact proof.

\section{System-spanning strings from the $J_1,J_3\rightarrow\infty$ constraints}
We have mentioned in the main text that the allowed configurations of the $J_1,J_3\rightarrow\infty$ limit of the Kagome Ising antiferromagnet are associated to the trajectories of three kinds of system-spanning strings:
\begin{itemize}
    \item \sA{} strings, drawn by connecting the centers of crosses arrows where arrow directions converge. 
    \item \sB{} strings between pairs of \sA{} strings, drawn by connecting the centers of hexagons where arrow directions diverge.
    \item \sC{} strings, consisting of Double Domain Walls (DDW's) with internal excitations (changes in arrow directions) and decorating \sB{} lines. 
\end{itemize}
In this section, we first illustrate how the $J_1,J_3\rightarrow\infty$ limit leads to local constraints, and then how those constraints lead to \sA{} and \sC{} strings which must be system-spanning.\par

Let us start from the $J_1 \rightarrow \infty$ limit. Because $J_2$ and $J_3$ are not competing with $J_1$, this limit imposes a simple 2-up 1-down, 2-down 1-up  (UUD-DDU) rule on every nearest-neighbor triangle of the kagome lattice. As illustrated in the End Matter (Fig. EM2), this constraint is equivalent to having exactly one dimer touching the center of every triangle. 
The limit $J_3 \rightarrow \infty$ plays a similar role, imposing the UUD-DDU rule on three independent triangular sublattices of the model. On the level of a kagome star, these two constraints conspire to give rise to all the configurations illustrated in Fig. EM1.
These configurations can be understood as local tiles or dominoes, which overlap and impose constraints, and must be matched to realize configurations of the kagome lattice model~\cite{vanhecke_solving_2021,colbois_partial_2022}.
We now show how these constraints give rise to system spanning strings. \par

\begin{figure}
    \centering
    \includegraphics[width=\linewidth]{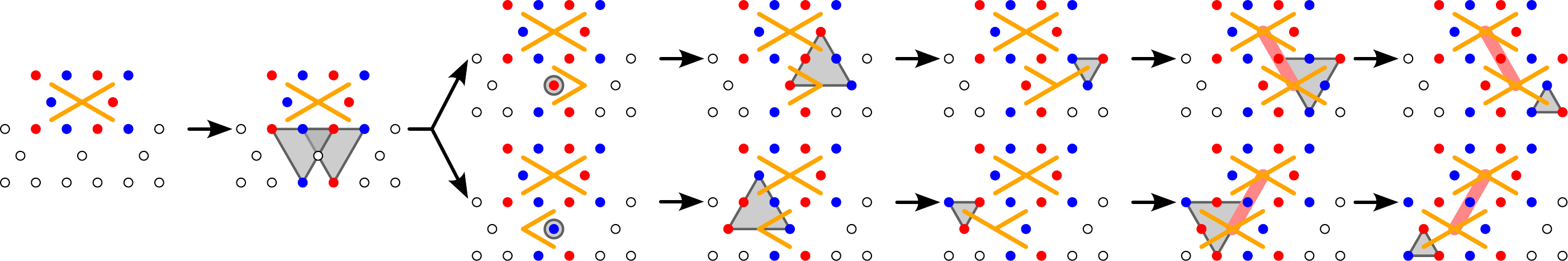}
    \caption{Starting from a hexagon with a cross, the constraints imposed by the $J_1,J_3\rightarrow\infty$ force the existence of another cross in one of the two adjacent hexagons beneath (by symmetry also above) it. At each step of the illustration the value of a spin is determined by the existence of two identical spins in a $J_1$ or $J_3$ triangle (highlighted in gray). The value of one of the spins may be freely chosen, leading to two branches for the diagram. This process may be continued indefinitely, leading to a system-spanning \sA{} strings which must not intersect.}
    \label{fig:continue_A_string}
\end{figure}

We begin with the case of the \sA{} strings. Our starting tile is a single crossed hexagon, and we seek to determine what are the allowed values for the surrounding spins. Since the energy of every $J_1$ and $J_3$ triangle is minimized, whenever two spins which interact with $J_1$ or $J_3$ are ferromagnetically aligned the value of the third spin on the triangle is set by constraints. Repeatedly applying this principle as shown in FIG. \ref{fig:continue_A_string}, we conclude that beneath (and above, by inversion symmetry) a crossed hexagon there must be another crossed hexagon. Successive application of this reasoning leads to the conclusion that crossed hexagons are always a part of system-spanning strings (see also the proof in Appendix of ~\cite{colbois_partial_2022}). A further consequence of this is that in the constrained system all crosses must have the same orientation: crosses with different orientations will propagate on average in different directions and must eventually intersect, which is not allowed.

We now move onto the case of the \sC{} strings.  We prove that if there is at least one \sA{} string in the system, and that it has the orientation shown in FIG. \ref{fig:continue_A_string}, then \sC{} strings must be oriented and non-crossing. We first prove some statements about \sC{}-strings and dimers, before restarting from the $J_3 \rightarrow\infty$ constraint to conclude the proof. 

The presence of one \sC{} string can be detected by the existence of a ferromagnetic horizontal bond (vertical dimer). Therefore, in order to prove that \sC{} strings cannot terminate it suffices to show that beneath (or above) a vertical dimer there must be another vertical dimer. We will need a useful intermediate result, which we now derive: adjacent vertical dimers must be between two crossed hexagons. The derivation is shown in FIG. 
\ref{fig:consecutive_dimers} and results from the successive application of the minimization of $J_1$ and $J_3$ interactions in different triangles. Using the assumption about the orientation of crosses in \sA{} strings, this means that consecutive vertical dimers are not allowed. \par
\begin{figure}
    \centering
    \includegraphics[width=\linewidth]{FIGSM9.pdf}
    \caption{In order to respect the $J_1,J_3\rightarrow\infty$ constraints, two consecutive ferromagnetic bonds along a row must be between two diagonally oriented crosses.}
    \label{fig:consecutive_dimers}
\end{figure}

Now, recall that the constraints imposed by infinite $J_3$ in one of the two triangular sub-lattices involving the dense rows force the spins in this sub-lattice to be ground states of the first-neighbor triangular Ising antiferromagnet. Recall as well that those configurations can be mapped onto the trajectories of system-spanning strings which separate stripe-ordered domains \cite{nourhani_communicating_2018,smerald_spin-liquid_2018,smerald_topological_2016}. An example of an allowed configuration on this sub-lattice is represented in FIG. \ref{fig:continue_C_string}. \par

Since, as shown before, consecutive vertical dimers are not allowed, there is no freedom in the value of spins on a second $J_3$ sub-lattice, except when they overlap with the string trajectories. Spins along the length of a string are not fully fixed by constraints, but regardless of the choice that is taken there will always be one vertical dimer adjacent to the string, which shows that strings of vertical dimers must never terminate. The remaining freedom of the spins on the strings is due to the internal freedom of the \sC{} strings.   

\begin{figure}
    \centering
    \includegraphics[width=\linewidth]{FIGSM10.pdf}
    \caption{Illustration of how the $J_1,J_3\rightarrow\infty$ constraints lead to system-spanning \sC{} strings. a) The allowed configurations of one of the $J_3$ sub-lattices may be described by trajectories of system-spanning strings (shown as dark lines), as in the first-neighbor triangular Ising antiferromagnet \cite{yokoi_dimer_1986,nourhani_communicating_2018}. b) In the presence of at least one horizontal cross (oriented as in Fig. \ref{fig:continue_A_string}), consecutive vertical dimers are not allowed. Therefore, the spins on the other $J_3$ sub-lattice of dense rows are also fully determined except on top of the strings. c) There are several possible choices for the spins along the black string (highlighted with a yellow star). Once the value of those spins has been chosen, the position of the vertical dimers is determined. It would also be possible to draw the black string using a different sub-lattice choice, which would have lead to the gray string. The strings drawn with different sublattice choices are always separated by half of a lattice spacing.  d) Applying the $J_1\rightarrow\infty$ rules, the presence of the vertical dimers leads to system-spanning \sC{} strings. In some cases the trajectory of the \sC{} string is not fully determined from the position of the vertical dimers, due to the remaining freedom in the spins highlighted with a yellow star.  }
    \label{fig:continue_C_string}
\end{figure}

\end{document}